\documentclass[aps, prl, twocolumn, superscriptaddress]{revtex4}
\usepackage{graphicx}
\usepackage{dcolumn}
\usepackage{bm}
\usepackage{natbib}

\begin{document}

\title{Metamaterials proposed as perfect magnetoelectrics}

\author{A. M. Shuvaev}
\author{S. Engelbrecht}
\author{M. Wunderlich}
\affiliation{Experimentelle Physik IV, Universit\"{a}t W\"{u}rzburg,
97074 W\"{u}rzburg, Germany} %
\author{A. Schneider}
\affiliation{Experimentelle Physik IV, Universit\"{a}t W\"{u}rzburg,
97074 W\"{u}rzburg, Germany} %
\affiliation{Fachbereich 1, Universit\"{a}t Bremen,  28359 Bremen, Germany} %
\author{A. Pimenov}
\affiliation{Experimentelle Physik IV, Universit\"{a}t W\"{u}rzburg,
97074 W\"{u}rzburg, Germany} %

\date{\today}

\begin{abstract}

Magnetoelectric susceptibility of a metamaterial built from split
ring resonators have been investigated both experimentally and
within an equivalent circuit model. The absolute values have been
shown to exceed by two orders of magnitude that of classical
magnetoelectric materials. The metamaterial investigated reaches the
theoretically predicted value of the magnetoelectric susceptibility
which is equal to the geometric average of the electric and magnetic
susceptibilities.

\end{abstract}


\maketitle

Magnetoelectric effect manifests a connection between electricity
and magnetism and has been predicted already 1860 by Pierre Curie,
but could be confirmed experimentally only about half a century ago
\cite{dell_book,fiebig_jpd_2005}. Up to now the experimentally
observed effects remain rather weak. The weakness of the
magnetoelectric susceptibility can to some extent be explained by
the absence of strong mechanisms which couple magnetism and
electricity on the microscopic level. As has been shown
theoretically \cite{brown_pr_1968, dell_book}, the allowed limiting
value of the magnetoelectric susceptibility ($\xi$) is quite large
and equals the geometric average of electric ($\chi_e$) and magnetic
($\chi_e$) susceptibilities:
\begin{equation}\label{me}
\xi^2 \leq \chi_e \chi_m
\end{equation}
In classical magnetoelectric materials like Cr$_2$O$_3$ the limiting
value of Eq. (\ref{me}) is failed by about two orders of magnitude
\cite{brown_pr_1968,fiebig_jpd_2005}. In efforts to increase the
value of the magnetoelectric effect, materials revealing both strong
electric and magnetic susceptibilities have been brought into
consideration. Especially close to phase transitions, the electric
and magnetic susceptibilities may diverge in ferroelectrics and
ferromagnets. Materials simultaneously showing the ferroelectricity
and ferromagnetism are called multiferroics and they are presently
the subject of intensive research
\cite{cheong_nm_2007,fiebig_jpd_2005,eerenstein_nature_2006}.

Assuming that the equality in Eq. (\ref{me}) holds, it can be
rewritten in the form $\xi= \sqrt{\chi_e \chi_m}$ and the
constitutive relationships
\begin{eqnarray}
M&=&\chi_m H + \chi_{me} E   \label{eq2} \\
P&=& \chi_e E + \chi_{em} H  \qquad , \label{eq3}
\end{eqnarray}
can be reduced to
\begin{equation}\label{eq4}
    M=iP \sqrt{\frac{\chi_m}{\chi_e}} \qquad ,
\end{equation}
i.e. electric polarization and magnetic moments must be directly
proportional to each other. Here we have rewritten $\chi_{em}= i
\xi$, and utilized the symmetry of the magnetoelectric coefficients
$\chi_{me}=-\chi_{em}$ \cite{dell_book,dell_philmag_1963}. The
condition in Eq. (\ref{eq4}) has been assumed in the first molecular
theories for magnetoelectric effect~\cite{dell_book} nearly a
century ago.

We note that within the arguments of thermodynamic stability
\cite{dell_philmag_1963} another magnetoelectric inequality can be
derived: $\xi^2 \leq \varepsilon \mu$ which utilizes the magnetic
permeability $\mu=1+\chi_m$ and electric permittivity
$\varepsilon=1+\chi_e$ instead of susceptibilities in Eq.
(\ref{me}).

Due to their unique electrodynamic properties, metamaterials (i.e.
artificial materials)~\cite{smith_science_2004, padilla_mat_2006,
shalaev_ol_2005} may show new ways to solve the problem of the
weakness of magnetoelectric coupling. Some examples of breakthroughs
in various topics of modern electromagnetism which were stimulated
by metamaterials are: reversing the laws of conventional
optics~\cite{pendry_today_2004},
cloaking~\cite{leonhardt_science_2006, pendry_science_2006}, or
overcoming the resolution limit of optical devices
\cite{pendry_prl_2000, grbic_prl_2004}. Mixing of electric and
magnetic responses is another useful property of metamaterials which
can be utilized to generate new effects. As has been shown recently,
the magnetoelectric coupling in metamaterials lead to strong optical
activity~\cite{tretyakov_book_2001, gonokami_prl_2005,
plum_apl_2007, thiel_adma_2007, liu_nphoton_2009} which is directly
connected to intrinsic chirality. In agreement with the theoretical
prediction \cite{pendry_science_2004, tretyakov_jewa_2003} the
chirality of metamaterials can lead to giant polarization rotation
and provide another routes to obtain negative refraction
\cite{monzon_prl_2005, zhang_prl_2009, zhou_prb_2009, plum_prb_2009}
for circularly polarized waves. Recently, for metamaterials with
zero permittivity and permeability an estimate for the limiting
value of the magnetoelectric effect $Re(\xi) \leq
Im{\sqrt{\varepsilon \mu}}$ has been obtained
\cite{tretyakov_jewa_2003}.

Many designs of metamaterials are based on split ring
resonators~\cite{pendry_mtt_1999}. These elements can be seen as the
smallest possible representations of the well known LC-circuit with
a single inductance loop as given by the metallic ring and a tiny
capacitance produced by the gap in the ring
\cite{linden_science_2004}. Split ring resonators have been
originally developed to achieve negative magnetic permeability which
is a key property for design of metamaterials with negative
refraction \cite{smith_science_2004}. As has been realized recently
\cite{marques_prb_2002}, the split ring resonators strongly modify
the interactions with electromagnetic radiation by introducing a so
called bianisotropy term into the set of basic equations
\cite{chen_pre_2005}, which is closely connected to the
magnetoelectric effect~\cite{eerenstein_nature_2006,
fiebig_jpd_2005}. This additional term in the constitutive relations
cross-couples the magnetic and electric fields within a split ring
resonator \cite{katsarakis_apl_2004, smith_jap_2006}. The
bianisotropy offers another degree of freedom
\cite{schneider_prl_2009} in controlling the properties of light.
Here we note that, contrary to chiral metamaterials, the
bianisotropy does not automatically lead to polarization rotation
for geometries parallel to the principal optical axes. In order to
obtain polarization rotation, these structures must be tilt or
measured within off-axis geometry. The corresponding effects
\cite{plum_apl_2008, plum_prl_2009} have been termed extrinsic
chirality.

In this work we show that metamaterials built from split ring
resonators achieve magnetoelectric effects equal to the
theoretically limiting value in Eq. (\ref{me}). To prove this we
investigate a metamaterial of split ring resonators within different
geometries, especially including those sensitive to
magnetoelectricity. This allowed to obtain electric, magnetic and
magnetoelectric susceptibilities and compare them with a simple
circuit model. We show that the metamaterial investigated indeed
reaches the theoretical limit for magnetoelectric coupling. This
value is due to direct proportionality of electric and magnetic
moments in split ring resonators.

Transmittance experiments at millimeter-wave frequencies (60 GHz
$<\nu<$ 120 GHz) were carried out in a Mach-Zehnder interferometer
arrangement \cite{kozlov_book,pimenov_prb_2005}. This arrangement
allows to measure both the intensity and the phase shift of the
radiation transmitted through the sample within controlled
polarization geometries. The split ring resonator arrays used in
present experiments were prepared by chemical etching of
copper-laminated board. The rings are typically 0.35 mm $\times$
0.35 mm in size with the gap width $d = 0.17$ mm. The lattice
constant of the metamaterial is $l = 0.7$ mm. The characteristic
parameters of various sets of the split ring resonators have been
varied within approximately a factor of two and showed qualitatively
similar results. Woven glass with a thickness of 0.56 mm was used as
a non-conductive substrate. The refractive index of the substrate
has been determined in a separate experiment as $n_s=2.07+0.04i$.

Split ring resonators seem to represent an ideal magnetoelectric
material fulfilling the condition $ \chi_{me} \chi_{em} = \chi_{e}
\chi_{m} $. Indeed, from the effective RLC-circuit model and simple
calculations we get
\begin{equation}\label{e1}
    \chi_e = nC  \cdot d^2 \cdot F(\omega) \qquad
    \chi_m = nC  \cdot S^2 \frac{\omega^2}{c^2}  \cdot F(\omega)
\end{equation}
\begin{equation}\label{e2}
    \chi_{em} = -\chi_{me} =
    nC  \cdot d \, S \frac{i \omega}{c}  \cdot F(\omega)
\end{equation}
Here $n$ is the density of the rings, $d$ and $C$ are the effective
width and capacitance of the gap, $S$ is the area of the rings, and
$\omega$ is the angular frequency. We use the Lorentz substitution
$F(\omega)=\omega_0^2/(\omega_0^2-\omega^2-i\omega\gamma)$ where
$\omega_0$ and $\gamma$ are resonance frequency and width,
respectively. The symmetry of the magnetoelectric coefficients is
fulfilled automatically in this model.

\begin{figure}[]
\includegraphics[angle=270, width=0.9\linewidth, clip]{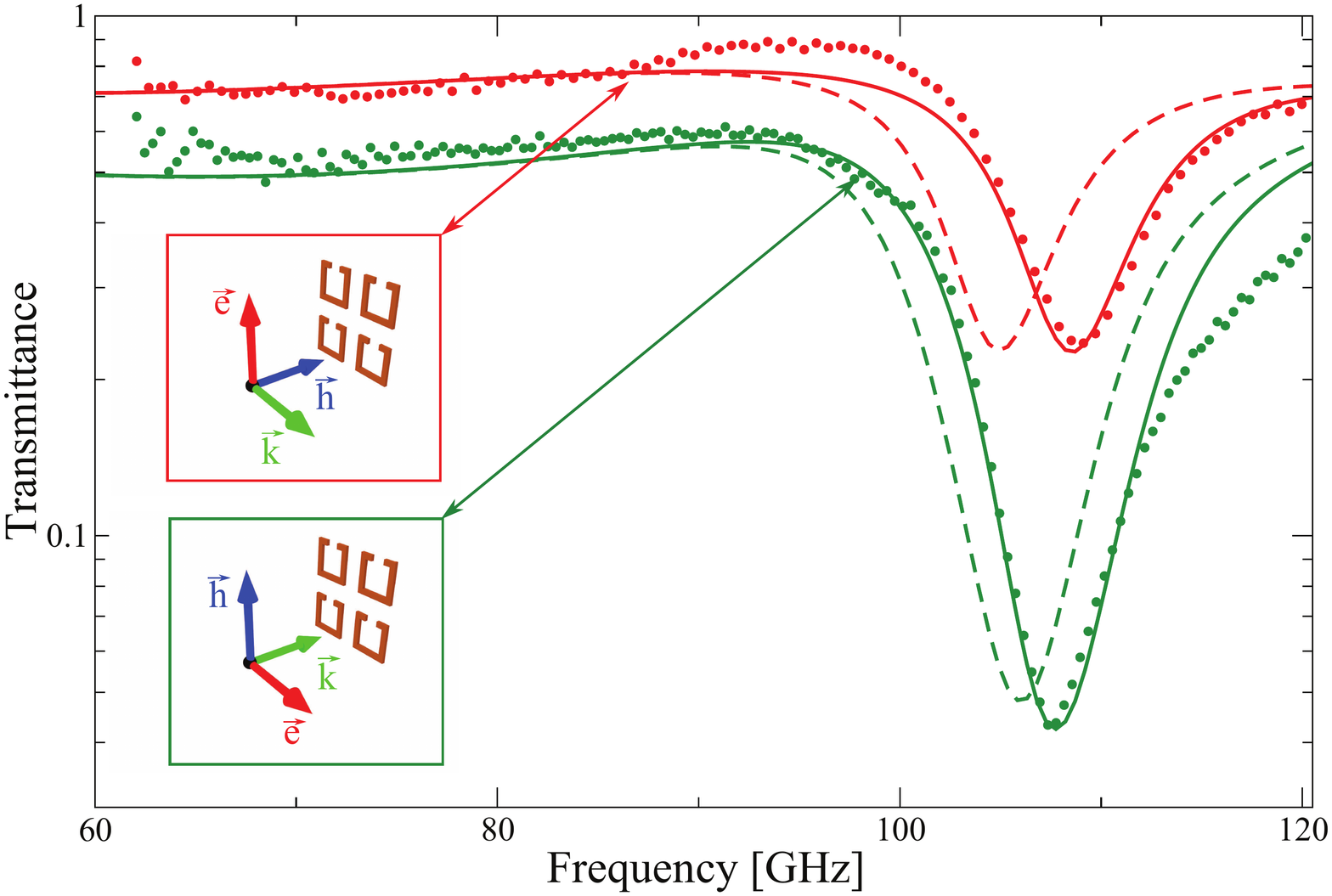}
\caption{Transmission characteristics of the metamaterial of split
ring resonators. Two experimental geometries shown are mostly
sensitive to the dielectric (green) and magnetic (red)
contributions. Symbols - experiment, lines are fits using Lorentzian
characteristics of the split rings. Solid lines - fits include
magnetoelectric coupling, dashed lines - magnetoelectric
susceptibility is set to zero, demonstrating negligible influence on
the spectra. The pictograms show the geometry of the experiments.}
\label{f90}
\end{figure}

In order to obtain the electrodynamic parameters of the split ring
resonators we have carried out the initial experiments within the
experimental geometries suggested in Ref. \cite{chen_pre_2005}
(shown in the insets to Fig. \ref{f90}). Three relevant geometries
in this case are "magnetic" ($\tilde{h}$ perpendicular to the plane
of the rings), "electric" ($\tilde{e}$ parallel to the gap of the
rings) and "magnetoelectric" (both excitations are realized
simultaneously). In these three geometries and within reasonable
approximation the effective refractive indexes basically determine
the transmittance close to the resonance and they are given by:
$n_e=\sqrt{\varepsilon-\xi^2/\mu}, \
n_m=\sqrt{\mu-\xi^2/\varepsilon}$, and $n_{me}=\sqrt{\varepsilon
\mu-\xi^2}$, respectively. Here we neglect the influence of the
substrate for simplicity. Although the magnetoelectric
susceptibility is included in these equations, the dominating terms
for typical parameters of the model are given by
$\sqrt{\varepsilon}, \ \sqrt{\mu}$, and $\sqrt{\varepsilon \mu}$. In
all cases the magnetoelectric susceptibility represents a weaker
correction under the square root. This is demonstrated in Fig.
\ref{f90} as we set the magnetoelectric susceptibility to zero,
which simply leads to a shift of the resonance frequency. In all
series of experiments with varying geometries the influence of the
magnetoelectric susceptibility was below the experimental accuracy.
This accuracy depends not only on experimental uncertainties, but
also on the assumptions of the circuit model, like neglecting of the
cross coupling effects, or assumption of infinitely small sizes of
the rings compared to the wavelength. On the contrary, electric and
magnetic geometries robustly depends on electric ($\chi_e$) and
magnetic ($\chi_m$) susceptibilities. Therefore, both
susceptibilities may be determined from the spectra in Fig.
\ref{f90}.

\begin{figure}[]
\includegraphics[angle=270, width=0.9\linewidth, clip]{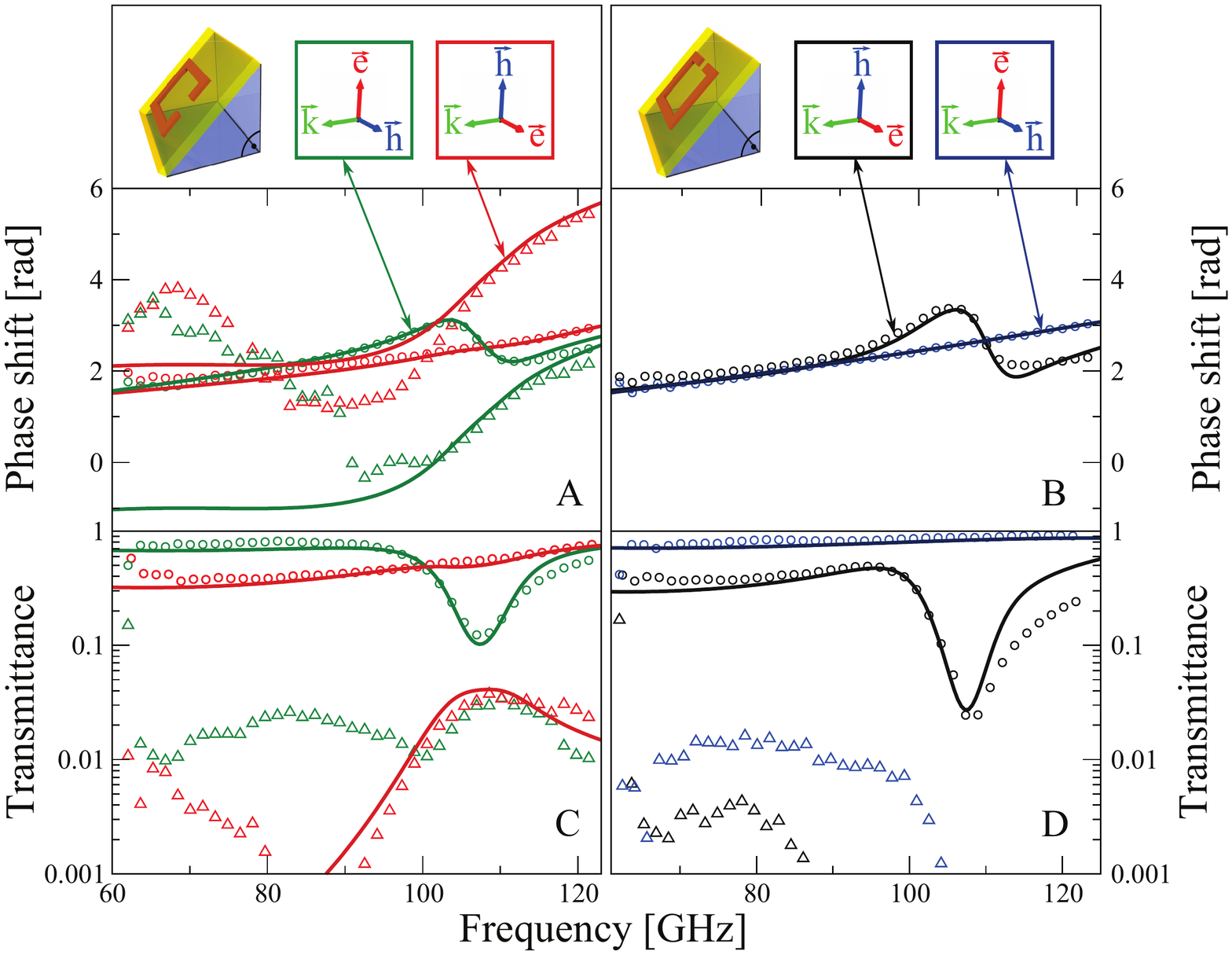}
\caption{Transmission and phase shift spectra of the metamaterial
within various tilted geometries as indicated. Symbols -
experimental data within the geometries as indicated, circles
correspond to the spectra with parallel polarizers, triangles - with
crossed polarizers. Solid lines are fits within the $4 \times 4$
matrix formalism \cite{berreman_josa_1972} and using Eq.
(\ref{e1},\ref{e2}) for susceptibilities curves. In the panel C the
model curves with crossed polarizers coincide for both geometries.}
\label{f45}
\end{figure}

The result of the experiments described above may now be  extended
to obtain the magnetoelectric susceptibility using further
geometries of the experiment. In order to get better sensitivity to
the magnetoelectric susceptibility, the sample must be measured in
the tilt geometry and the signals in parallel and in crossed
polarizers have to be compared (see inset in Fig. \ref{f45}). An
example of such tilt geometries has been presented in
Ref.~\cite{smith_jap_2006}, where the existence of a cross-coupling
terms has been detected. The effectiveness of the tilt experiments
can be demonstrated within the simplified assumptions. If we assume
$ka <<1$, then the corresponding Maxwell equations may be solved
more easily leading to analytical expressions for all relevant
geometries (e.g. Ref.~\cite{oksanen_jewa_1990}). Here $k=2
\pi/\lambda$ is the wavevector of the electromagnetic wave and $a$
is the sample thickness. If we further neglect the $(ka)^2$ terms
compared to the terms linear in $ka$, then e.g. for both geometries
in the left panels in Fig. \ref{f45} and within $45^{\circ}$
incidence the following expression for the complex transmittance in
crossed polarizers can be written:
\begin{equation}\label{esimpl}
t(\omega)=\frac{k a \cdot \xi}{2\sqrt{2}\mu+ik
a[1+\xi^2-\mu(\varepsilon+2)]}
\end{equation}
The last formula clearly shows that the transmittance in crossed
polarizers is directly proportional to the magnetoelectric
susceptibility. Therefore, this geometry provides a sensitive tool
for magnetoelectric effect. In the following, the calculations have
been done within the exact $4 \times 4$ matrix formalism as
described in Ref. \cite{berreman_josa_1972}.

Figure \ref{f45} shows transmittance and phase shift spectra in four
most important tilt geometries, both in parallel and in crossed
polarizers.  The main result of this data is given in the left
bottom panel of Fig. \ref{f45} showing the nonzero transmittance in
crossed polarizers close to the resonance of the rings ($109$ GHz).
As expected already from the simplified equation Eq. (\ref{esimpl}),
reasonably strong signal can be observed in geometries with crossed
polarizers (red and green triangles). In the left bottom panel of
Fig. \ref{f45} the nonzero signal below about 90 GHz for crossed
geometry are due to the radiation leakage around the sample and
should be neglected. The leaky signal can be also seen in the phase
shifts data as shown in the left upper panel. Below about 90 GHz the
amplitude of the signal is getting low and the phase measuring
system looses the stability (red and green triangles).

We note another interesting feature of Fig. \ref{f45}. In the
geometry with ac electric fields within the plane of the rings and
parallel to the gap (black circles in the right panels of Fig.
\ref{f45}) the resonance is clearly seen in parallel polarizers.
However, we observe no signal in crossed polarizers in this geometry
(black triangles). In two geometries presented in the right panels
of Fig. \ref{f45} solely a stray signal could be detected. This
result is supported both by simple and rigorous theories. Both
calculation methods predict zero amplitude of the perpendicular
polarization and reflect the symmetry of the problem.

\begin{figure}[]
\includegraphics[width=0.45\linewidth,clip]{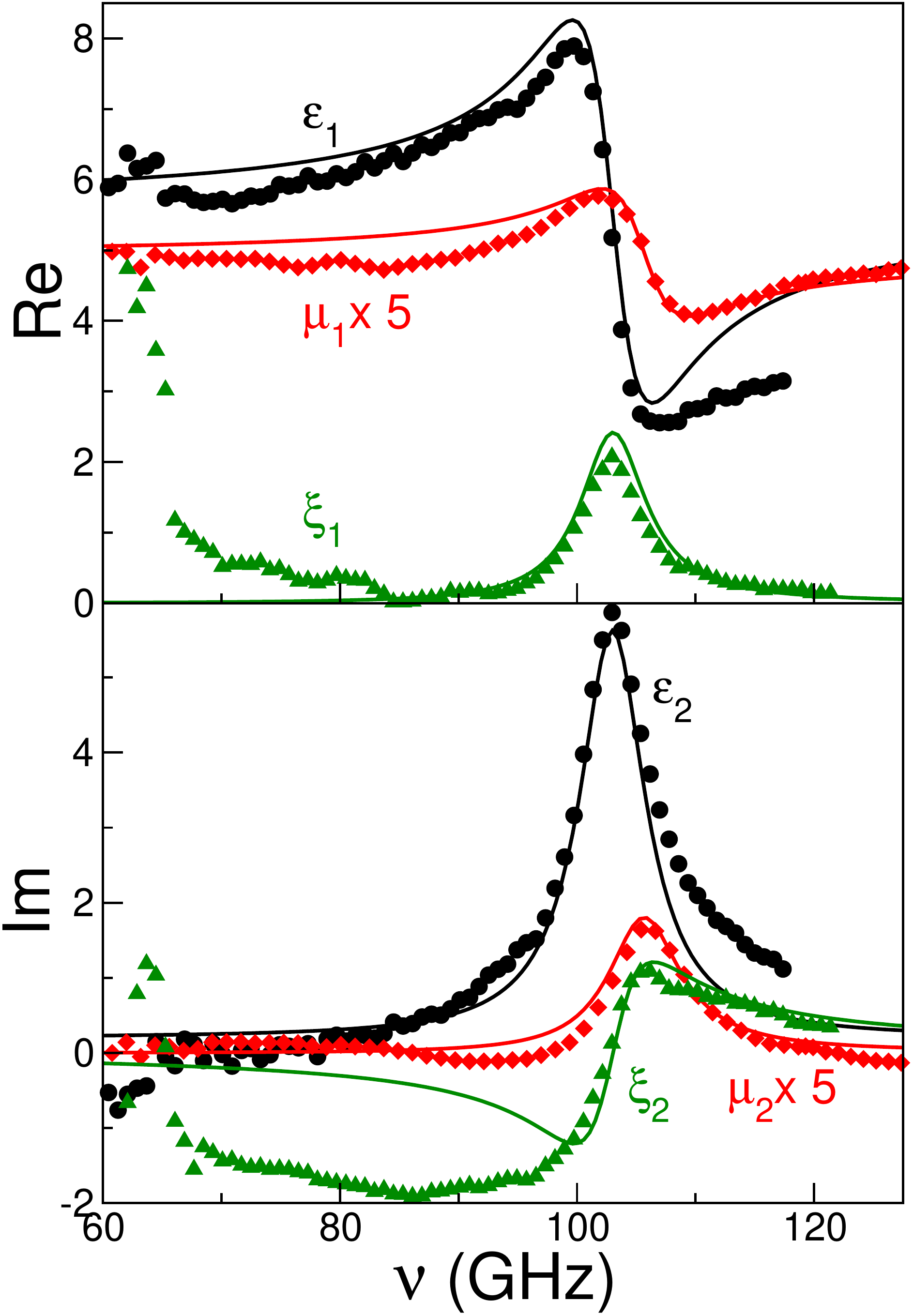}
\caption{Permittivity, permeability and bianisotropy of the
metamaterial used in the present work. Symbols - experiment, lines -
circuit model given in Eqs. (\ref{e1}, \ref{e2}). The
magnetoelectric susceptibility has been assumed to obey
$\xi/\sqrt{\chi_{e} \chi_{m}} = 0.9 \pm 0.15$ instead of Eq.
(\ref{me}).} \label{ksi}
\end{figure}

The results of the transmittance experiments in perpendicular (Fig.
\ref{f90}) and tilted (Fig. \ref{f45}) orientations are sufficient
to determine all three complex susceptibilities for the metamaterial
of split ring resonators. The results of the self-consistent
calculation of these parameters based on the $4 \times 4$ matrix
formalism is presented in Fig. \ref{ksi}. As could be already
expected from the fit of the transmittance spectra, all three
susceptibilities reveal a resonance-like form. The results in Fig.
\ref{ksi} clearly demonstrate that the metamaterial of split ring
resonators indeed models a perfect magnetoelectric with
$\xi^2=\chi_e \chi_m$. Here solely a factor of $0.9 \pm 0.15$ had to
be introduced in order to obtain the self-consistence of the data.
Taking into account the assumptions made, this factor most probably
reflects the uncertainties of the experiment. Finally, we recall
that the condition $\xi^2=\chi_e \chi_m$ is expected for all
materials revealing direct proportionality between electric and
magnetic moments ($M=\alpha P$). Here  $\alpha$ is a material
constant which equals $\alpha =iS \omega/cd$ for split ring
resonators.

In summary, using millimeter wave spectroscopy of the complex
transmission coefficient the electrodynamic properties of a
metamaterial of split ring resonators have been investigated. The
sensitivity to the magnetoelectric effect has been obtained within
tilt sample geometry and calculated within $4 \times 4$ matrix
formalism. We prove experimentally and within a circuit model
calculation that metamaterials from split ring resonators reach the
maximum theoretical values of the magnetoelectric susceptibility
limited by $\xi^2 \leq \chi_e \chi_m$. This value appears to be
about two orders of magnitude above the typical coupling constants
for conventional magnetoelectrics like Cr$_2$O$_3$.

\bibliography{rings_me}

\begin{thebibliography}{38}
\expandafter\ifx\csname natexlab\endcsname\relax\def\natexlab#1{#1}\fi
\expandafter\ifx\csname bibnamefont\endcsname\relax
  \def\bibnamefont#1{#1}\fi
\expandafter\ifx\csname bibfnamefont\endcsname\relax
  \def\bibfnamefont#1{#1}\fi
\expandafter\ifx\csname citenamefont\endcsname\relax
  \def\citenamefont#1{#1}\fi
\expandafter\ifx\csname url\endcsname\relax
  \def\url#1{\texttt{#1}}\fi
\expandafter\ifx\csname urlprefix\endcsname\relax\def\urlprefix{URL }\fi
\providecommand{\bibinfo}[2]{#2}
\providecommand{\eprint}[2][]{\url{#2}}

\bibitem[{\citenamefont{O'Dell}(1970)}]{dell_book}
\bibinfo{author}{\bibfnamefont{T.~H.} \bibnamefont{O'Dell}},
  \emph{\bibinfo{title}{The electrodynamics of magneto-electric media}}
  (\bibinfo{publisher}{North-Holland Publishing}, \bibinfo{address}{Amsterdam},
  \bibinfo{year}{1970}).

\bibitem[{\citenamefont{Fiebig}(2005)}]{fiebig_jpd_2005}
\bibinfo{author}{\bibfnamefont{M.}~\bibnamefont{Fiebig}},
  \bibinfo{journal}{Journal of Physics D: Applied Physics}
  \textbf{\bibinfo{volume}{38}}, \bibinfo{pages}{R123} (\bibinfo{year}{2005}),
  \urlprefix\url{http://stacks.iop.org/0022-3727/38/R123}.

\bibitem[{\citenamefont{Brown et~al.}(1968)\citenamefont{Brown, Hornreich, and
  Shtrikman}}]{brown_pr_1968}
\bibinfo{author}{\bibfnamefont{W.~F.} \bibnamefont{Brown}},
  \bibinfo{author}{\bibfnamefont{R.~M.} \bibnamefont{Hornreich}},
  \bibnamefont{and}
  \bibinfo{author}{\bibfnamefont{S.}~\bibnamefont{Shtrikman}},
  \bibinfo{journal}{Phys. Rev.} \textbf{\bibinfo{volume}{168}},
  \bibinfo{pages}{574} (\bibinfo{year}{1968}).

\bibitem[{\citenamefont{Cheong and Mostovoy}(2007)}]{cheong_nm_2007}
\bibinfo{author}{\bibfnamefont{S.-W.} \bibnamefont{Cheong}} \bibnamefont{and}
  \bibinfo{author}{\bibfnamefont{M.}~\bibnamefont{Mostovoy}},
  \bibinfo{journal}{NATURE MATERIALS} \textbf{\bibinfo{volume}{6}},
  \bibinfo{pages}{13} (\bibinfo{year}{2007}), ISSN \bibinfo{issn}{1476-1122}.

\bibitem[{\citenamefont{Eerenstein et~al.}(2006)\citenamefont{Eerenstein,
  Mathur, and Scott}}]{eerenstein_nature_2006}
\bibinfo{author}{\bibfnamefont{W.}~\bibnamefont{Eerenstein}},
  \bibinfo{author}{\bibfnamefont{N.~D.} \bibnamefont{Mathur}},
  \bibnamefont{and} \bibinfo{author}{\bibfnamefont{J.~F.} \bibnamefont{Scott}},
  \bibinfo{journal}{Nature} \textbf{\bibinfo{volume}{442}},
  \bibinfo{pages}{759} (\bibinfo{year}{2006}), ISSN \bibinfo{issn}{0028-0836},
  \urlprefix\url{http://dx.doi.org/10.1038/nature05023}.

\bibitem[{\citenamefont{O~Dell}(1963)}]{dell_philmag_1963}
\bibinfo{author}{\bibfnamefont{T.~H.} \bibnamefont{O~Dell}},
  \bibinfo{journal}{Phil. Mag.} \textbf{\bibinfo{volume}{8}},
  \bibinfo{pages}{411} (\bibinfo{year}{1963}), ISSN \bibinfo{issn}{0031-8086}.

\bibitem[{\citenamefont{Smith et~al.}(2004)\citenamefont{Smith, Pendry, and
  Wiltshire}}]{smith_science_2004}
\bibinfo{author}{\bibfnamefont{D.~R.} \bibnamefont{Smith}},
  \bibinfo{author}{\bibfnamefont{J.~B.} \bibnamefont{Pendry}},
  \bibnamefont{and} \bibinfo{author}{\bibfnamefont{M.~C.~K.}
  \bibnamefont{Wiltshire}}, \bibinfo{journal}{Science}
  \textbf{\bibinfo{volume}{305}}, \bibinfo{pages}{788} (\bibinfo{year}{2004}),
  \eprint{http://www.sciencemag.org/cgi/reprint/305/5685/788.pdf},
  \urlprefix\url{http://www.sciencemag.org/cgi/content/abstract/305/5685/788}.

\bibitem[{\citenamefont{Padilla et~al.}(2006)\citenamefont{Padilla, Basov, and
  Smith}}]{padilla_mat_2006}
\bibinfo{author}{\bibfnamefont{W.~J.} \bibnamefont{Padilla}},
  \bibinfo{author}{\bibfnamefont{D.~N.} \bibnamefont{Basov}}, \bibnamefont{and}
  \bibinfo{author}{\bibfnamefont{D.~R.} \bibnamefont{Smith}},
  \bibinfo{journal}{Materials Today} \textbf{\bibinfo{volume}{9}},
  \bibinfo{pages}{28} (\bibinfo{year}{2006}), ISSN \bibinfo{issn}{1369-7021},
  \urlprefix\url{http://www.sciencedirect.com/science/article/B6X1J-4K8Y762-P/%
2/f3f6e6c7b9f12f09e3c505c1793e9881}.

\bibitem[{\citenamefont{Shalaev et~al.}(2005)\citenamefont{Shalaev, Cai,
  Chettiar, Yuan, Sarychev, Drachev, and Kildishev}}]{shalaev_ol_2005}
\bibinfo{author}{\bibfnamefont{V.~M.} \bibnamefont{Shalaev}},
  \bibinfo{author}{\bibfnamefont{W.}~\bibnamefont{Cai}},
  \bibinfo{author}{\bibfnamefont{U.~K.} \bibnamefont{Chettiar}},
  \bibinfo{author}{\bibfnamefont{H.-K.} \bibnamefont{Yuan}},
  \bibinfo{author}{\bibfnamefont{A.~K.} \bibnamefont{Sarychev}},
  \bibinfo{author}{\bibfnamefont{V.~P.} \bibnamefont{Drachev}},
  \bibnamefont{and} \bibinfo{author}{\bibfnamefont{A.~V.}
  \bibnamefont{Kildishev}}, \bibinfo{journal}{Opt. Lett.}
  \textbf{\bibinfo{volume}{30}}, \bibinfo{pages}{3356} (\bibinfo{year}{2005}),
  \urlprefix\url{http://ol.osa.org/abstract.cfm?URI=ol-30-24-3356}.

\bibitem[{\citenamefont{Pendry and Smith}(2004)}]{pendry_today_2004}
\bibinfo{author}{\bibfnamefont{J.~B.} \bibnamefont{Pendry}} \bibnamefont{and}
  \bibinfo{author}{\bibfnamefont{D.~R.} \bibnamefont{Smith}},
  \bibinfo{journal}{Physics Today} \textbf{\bibinfo{volume}{57}},
  \bibinfo{pages}{37} (\bibinfo{year}{2004}),
  \urlprefix\url{http://link.aip.org/link/?PTO/57/37/1}.

\bibitem[{\citenamefont{Leonhardt}(2006)}]{leonhardt_science_2006}
\bibinfo{author}{\bibfnamefont{U.}~\bibnamefont{Leonhardt}},
  \bibinfo{journal}{Science} \textbf{\bibinfo{volume}{312}},
  \bibinfo{pages}{1777} (\bibinfo{year}{2006}),
  \eprint{http://www.sciencemag.org/cgi/reprint/312/5781/1777.pdf},
  \urlprefix\url{http://www.sciencemag.org/cgi/content/abstract/312/5781/1777}.

\bibitem[{\citenamefont{Pendry et~al.}(2006)\citenamefont{Pendry, Schurig, and
  Smith}}]{pendry_science_2006}
\bibinfo{author}{\bibfnamefont{J.~B.} \bibnamefont{Pendry}},
  \bibinfo{author}{\bibfnamefont{D.}~\bibnamefont{Schurig}}, \bibnamefont{and}
  \bibinfo{author}{\bibfnamefont{D.~R.} \bibnamefont{Smith}},
  \bibinfo{journal}{Science} \textbf{\bibinfo{volume}{312}},
  \bibinfo{pages}{1780} (\bibinfo{year}{2006}),
  \eprint{http://www.sciencemag.org/cgi/reprint/312/5781/1780.pdf},
  \urlprefix\url{http://www.sciencemag.org/cgi/content/abstract/312/5781/1780}.

\bibitem[{\citenamefont{Pendry}(2000)}]{pendry_prl_2000}
\bibinfo{author}{\bibfnamefont{J.~B.} \bibnamefont{Pendry}},
  \bibinfo{journal}{Phys. Rev. Lett.} \textbf{\bibinfo{volume}{85}},
  \bibinfo{pages}{3966} (\bibinfo{year}{2000}).

\bibitem[{\citenamefont{Grbic and Eleftheriades}(2004)}]{grbic_prl_2004}
\bibinfo{author}{\bibfnamefont{A.}~\bibnamefont{Grbic}} \bibnamefont{and}
  \bibinfo{author}{\bibfnamefont{G.~V.} \bibnamefont{Eleftheriades}},
  \bibinfo{journal}{Phys. Rev. Lett.} \textbf{\bibinfo{volume}{92}},
  \bibinfo{pages}{117403} (\bibinfo{year}{2004}).

\bibitem[{\citenamefont{Serdyukov et~al.}(2001)\citenamefont{Serdyukov,
  Semchenko, Tretyakov, and Sihvola}}]{tretyakov_book_2001}
\bibinfo{author}{\bibfnamefont{A.}~\bibnamefont{Serdyukov}},
  \bibinfo{author}{\bibfnamefont{I.}~\bibnamefont{Semchenko}},
  \bibinfo{author}{\bibfnamefont{S.}~\bibnamefont{Tretyakov}},
  \bibnamefont{and} \bibinfo{author}{\bibfnamefont{A.}~\bibnamefont{Sihvola}},
  \emph{\bibinfo{title}{Electromagnetics of bi-anisotropic materials: Theory
  and applications}} (\bibinfo{publisher}{Gordon and Breach},
  \bibinfo{address}{Amsterdam}, \bibinfo{year}{2001}), \bibinfo{edition}{1st}
  ed., ISBN \bibinfo{isbn}{9056993275},
  \urlprefix\url{http://www.amazon.com/gp/product/9056993275}.

\bibitem[{\citenamefont{Kuwata-Gonokami
  et~al.}(2005)\citenamefont{Kuwata-Gonokami, Saito, Ino, Kauranen, Jefimovs,
  Vallius, Turunen, and Svirko}}]{gonokami_prl_2005}
\bibinfo{author}{\bibfnamefont{M.}~\bibnamefont{Kuwata-Gonokami}},
  \bibinfo{author}{\bibfnamefont{N.}~\bibnamefont{Saito}},
  \bibinfo{author}{\bibfnamefont{Y.}~\bibnamefont{Ino}},
  \bibinfo{author}{\bibfnamefont{M.}~\bibnamefont{Kauranen}},
  \bibinfo{author}{\bibfnamefont{K.}~\bibnamefont{Jefimovs}},
  \bibinfo{author}{\bibfnamefont{T.}~\bibnamefont{Vallius}},
  \bibinfo{author}{\bibfnamefont{J.}~\bibnamefont{Turunen}}, \bibnamefont{and}
  \bibinfo{author}{\bibfnamefont{Y.}~\bibnamefont{Svirko}},
  \bibinfo{journal}{Phys. Rev. Lett.} \textbf{\bibinfo{volume}{95}},
  \bibinfo{pages}{227401} (\bibinfo{year}{2005}).

\bibitem[{\citenamefont{Plum et~al.}(2007)\citenamefont{Plum, Fedotov,
  Schwanecke, Zheludev, and Chen}}]{plum_apl_2007}
\bibinfo{author}{\bibfnamefont{E.}~\bibnamefont{Plum}},
  \bibinfo{author}{\bibfnamefont{V.~A.} \bibnamefont{Fedotov}},
  \bibinfo{author}{\bibfnamefont{A.~S.} \bibnamefont{Schwanecke}},
  \bibinfo{author}{\bibfnamefont{N.~I.} \bibnamefont{Zheludev}},
  \bibnamefont{and} \bibinfo{author}{\bibfnamefont{Y.}~\bibnamefont{Chen}},
  \bibinfo{journal}{Applied Physics Letters} \textbf{\bibinfo{volume}{90}},
  \bibinfo{eid}{223113} (pages~\bibinfo{numpages}{3}) (\bibinfo{year}{2007}),
  \urlprefix\url{http://link.aip.org/link/?APL/90/223113/1}.

\bibitem[{\citenamefont{Thiel et~al.}(2007)\citenamefont{Thiel, Decker, Deubel,
  Wegener, Linden, and von Freymann}}]{thiel_adma_2007}
\bibinfo{author}{\bibfnamefont{M.}~\bibnamefont{Thiel}},
  \bibinfo{author}{\bibfnamefont{M.}~\bibnamefont{Decker}},
  \bibinfo{author}{\bibfnamefont{M.}~\bibnamefont{Deubel}},
  \bibinfo{author}{\bibfnamefont{M.}~\bibnamefont{Wegener}},
  \bibinfo{author}{\bibfnamefont{S.}~\bibnamefont{Linden}}, \bibnamefont{and}
  \bibinfo{author}{\bibfnamefont{G.}~\bibnamefont{von Freymann}},
  \bibinfo{journal}{Advanced Materials} \textbf{\bibinfo{volume}{19}},
  \bibinfo{pages}{207} (\bibinfo{year}{2007}), ISSN \bibinfo{issn}{1521-4095},
  \urlprefix\url{http://dx.doi.org/10.1002/adma.200601497}.

\bibitem[{\citenamefont{Liu et~al.}(2009)\citenamefont{Liu, Liu, Zhu, and
  Giessen}}]{liu_nphoton_2009}
\bibinfo{author}{\bibfnamefont{N.}~\bibnamefont{Liu}},
  \bibinfo{author}{\bibfnamefont{H.}~\bibnamefont{Liu}},
  \bibinfo{author}{\bibfnamefont{S.}~\bibnamefont{Zhu}}, \bibnamefont{and}
  \bibinfo{author}{\bibfnamefont{H.}~\bibnamefont{Giessen}},
  \bibinfo{journal}{Nat Photon} \textbf{\bibinfo{volume}{3}},
  \bibinfo{pages}{157} (\bibinfo{year}{2009}), ISSN \bibinfo{issn}{1749-4885},
  \urlprefix\url{http://dx.doi.org/10.1038/nphoton.2009.4}.

\bibitem[{\citenamefont{Pendry}(2004)}]{pendry_science_2004}
\bibinfo{author}{\bibfnamefont{J.~B.} \bibnamefont{Pendry}},
  \bibinfo{journal}{Science} \textbf{\bibinfo{volume}{306}},
  \bibinfo{pages}{1353} (\bibinfo{year}{2004}),
  \eprint{http://www.sciencemag.org/cgi/reprint/306/5700/1353.pdf},
  \urlprefix\url{http://www.sciencemag.org/cgi/content/abstract/306/5700/1353}.

\bibitem[{\citenamefont{Tretyakov et~al.}(2003)\citenamefont{Tretyakov,
  Nefedov, Sihvola, Maslovski, and Simovski}}]{tretyakov_jewa_2003}
\bibinfo{author}{\bibfnamefont{S.}~\bibnamefont{Tretyakov}},
  \bibinfo{author}{\bibfnamefont{I.}~\bibnamefont{Nefedov}},
  \bibinfo{author}{\bibfnamefont{A.}~\bibnamefont{Sihvola}},
  \bibinfo{author}{\bibfnamefont{S.}~\bibnamefont{Maslovski}},
  \bibnamefont{and} \bibinfo{author}{\bibfnamefont{C.}~\bibnamefont{Simovski}},
  \bibinfo{journal}{Journal of Electromagnetic Waves and Applications}
  \textbf{\bibinfo{volume}{17}}, \bibinfo{pages}{695} (\bibinfo{year}{2003}),
  \urlprefix\url{http://brill.publisher.ingentaconnect.com/content/vsp/jew/200%
3/00000017/00000005/art00002}.

\bibitem[{\citenamefont{Monzon and Forester}(2005)}]{monzon_prl_2005}
\bibinfo{author}{\bibfnamefont{C.}~\bibnamefont{Monzon}} \bibnamefont{and}
  \bibinfo{author}{\bibfnamefont{D.~W.} \bibnamefont{Forester}},
  \bibinfo{journal}{Phys. Rev. Lett.} \textbf{\bibinfo{volume}{95}},
  \bibinfo{pages}{123904} (\bibinfo{year}{2005}).

\bibitem[{\citenamefont{Zhang et~al.}(2009)\citenamefont{Zhang, Park, Li, Lu,
  Zhang, and Zhang}}]{zhang_prl_2009}
\bibinfo{author}{\bibfnamefont{S.}~\bibnamefont{Zhang}},
  \bibinfo{author}{\bibfnamefont{Y.-S.} \bibnamefont{Park}},
  \bibinfo{author}{\bibfnamefont{J.}~\bibnamefont{Li}},
  \bibinfo{author}{\bibfnamefont{X.}~\bibnamefont{Lu}},
  \bibinfo{author}{\bibfnamefont{W.}~\bibnamefont{Zhang}}, \bibnamefont{and}
  \bibinfo{author}{\bibfnamefont{X.}~\bibnamefont{Zhang}},
  \bibinfo{journal}{Physical Review Letters} \textbf{\bibinfo{volume}{102}},
  \bibinfo{eid}{023901} (pages~\bibinfo{numpages}{4}) (\bibinfo{year}{2009}),
  \urlprefix\url{http://link.aps.org/abstract/PRL/v102/e023901}.

\bibitem[{\citenamefont{Zhou et~al.}(2009)\citenamefont{Zhou, Dong, Wang,
  Koschny, Kafesaki, and Soukoulis}}]{zhou_prb_2009}
\bibinfo{author}{\bibfnamefont{J.}~\bibnamefont{Zhou}},
  \bibinfo{author}{\bibfnamefont{J.}~\bibnamefont{Dong}},
  \bibinfo{author}{\bibfnamefont{B.}~\bibnamefont{Wang}},
  \bibinfo{author}{\bibfnamefont{T.}~\bibnamefont{Koschny}},
  \bibinfo{author}{\bibfnamefont{M.}~\bibnamefont{Kafesaki}}, \bibnamefont{and}
  \bibinfo{author}{\bibfnamefont{C.~M.} \bibnamefont{Soukoulis}},
  \bibinfo{journal}{Physical Review B (Condensed Matter and Materials Physics)}
  \textbf{\bibinfo{volume}{79}}, \bibinfo{eid}{121104}
  (pages~\bibinfo{numpages}{4}) (\bibinfo{year}{2009}),
  \urlprefix\url{http://link.aps.org/abstract/PRB/v79/e121104}.

\bibitem[{\citenamefont{Plum et~al.}(2009{\natexlab{a}})\citenamefont{Plum,
  Zhou, Dong, Fedotov, Koschny, Soukoulis, and Zheludev}}]{plum_prb_2009}
\bibinfo{author}{\bibfnamefont{E.}~\bibnamefont{Plum}},
  \bibinfo{author}{\bibfnamefont{J.}~\bibnamefont{Zhou}},
  \bibinfo{author}{\bibfnamefont{J.}~\bibnamefont{Dong}},
  \bibinfo{author}{\bibfnamefont{V.~A.} \bibnamefont{Fedotov}},
  \bibinfo{author}{\bibfnamefont{T.}~\bibnamefont{Koschny}},
  \bibinfo{author}{\bibfnamefont{C.~M.} \bibnamefont{Soukoulis}},
  \bibnamefont{and} \bibinfo{author}{\bibfnamefont{N.~I.}
  \bibnamefont{Zheludev}}, \bibinfo{journal}{Physical Review B (Condensed
  Matter and Materials Physics)} \textbf{\bibinfo{volume}{79}},
  \bibinfo{eid}{035407} (pages~\bibinfo{numpages}{6})
  (\bibinfo{year}{2009}{\natexlab{a}}),
  \urlprefix\url{http://link.aps.org/abstract/PRB/v79/e035407}.

\bibitem[{\citenamefont{Pendry et~al.}(1999)\citenamefont{Pendry, Holden,
  Robbins, and Stewart}}]{pendry_mtt_1999}
\bibinfo{author}{\bibfnamefont{J.~B.} \bibnamefont{Pendry}},
  \bibinfo{author}{\bibfnamefont{A.~J.} \bibnamefont{Holden}},
  \bibinfo{author}{\bibfnamefont{D.~J.} \bibnamefont{Robbins}},
  \bibnamefont{and} \bibinfo{author}{\bibfnamefont{W.~J.}
  \bibnamefont{Stewart}}, \bibinfo{journal}{Microwave Theory and Techniques,
  IEEE Transactions on} \textbf{\bibinfo{volume}{47}}, \bibinfo{pages}{2075}
  (\bibinfo{year}{1999}), ISSN \bibinfo{issn}{0018-9480},
  \urlprefix\url{http://ieeexplore.ieee.org/xpl/freeabs_all.jsp?isnumber=17309%
&arnumber=798002&count=18&index=2}.

\bibitem[{\citenamefont{Linden et~al.}(2004)\citenamefont{Linden, Enkrich,
  Wegener, Zhou, Koschny, and Soukoulis}}]{linden_science_2004}
\bibinfo{author}{\bibfnamefont{S.}~\bibnamefont{Linden}},
  \bibinfo{author}{\bibfnamefont{C.}~\bibnamefont{Enkrich}},
  \bibinfo{author}{\bibfnamefont{M.}~\bibnamefont{Wegener}},
  \bibinfo{author}{\bibfnamefont{J.}~\bibnamefont{Zhou}},
  \bibinfo{author}{\bibfnamefont{T.}~\bibnamefont{Koschny}}, \bibnamefont{and}
  \bibinfo{author}{\bibfnamefont{C.~M.} \bibnamefont{Soukoulis}},
  \bibinfo{journal}{Science} \textbf{\bibinfo{volume}{306}},
  \bibinfo{pages}{1351} (\bibinfo{year}{2004}),
  \eprint{http://www.sciencemag.org/cgi/reprint/306/5700/1351.pdf},
  \urlprefix\url{http://www.sciencemag.org/cgi/content/abstract/306/5700/1351}.

\bibitem[{\citenamefont{Marqu\'es et~al.}(2002)\citenamefont{Marqu\'es, Medina,
  and Rafii-El-Idrissi}}]{marques_prb_2002}
\bibinfo{author}{\bibfnamefont{R.}~\bibnamefont{Marqu\'es}},
  \bibinfo{author}{\bibfnamefont{F.}~\bibnamefont{Medina}}, \bibnamefont{and}
  \bibinfo{author}{\bibfnamefont{R.}~\bibnamefont{Rafii-El-Idrissi}},
  \bibinfo{journal}{Phys. Rev. B} \textbf{\bibinfo{volume}{65}},
  \bibinfo{pages}{144440} (\bibinfo{year}{2002}).

\bibitem[{\citenamefont{Chen et~al.}(2005)\citenamefont{Chen, Wu, Kong, and
  Grzegorczyk}}]{chen_pre_2005}
\bibinfo{author}{\bibfnamefont{X.}~\bibnamefont{Chen}},
  \bibinfo{author}{\bibfnamefont{B.-I.} \bibnamefont{Wu}},
  \bibinfo{author}{\bibfnamefont{J.~A.} \bibnamefont{Kong}}, \bibnamefont{and}
  \bibinfo{author}{\bibfnamefont{T.~M.} \bibnamefont{Grzegorczyk}},
  \bibinfo{journal}{Phys. Rev. E} \textbf{\bibinfo{volume}{71}},
  \bibinfo{pages}{046610} (\bibinfo{year}{2005}).

\bibitem[{\citenamefont{Katsarakis et~al.}(2004)\citenamefont{Katsarakis,
  Koschny, Kafesaki, Economou, and Soukoulis}}]{katsarakis_apl_2004}
\bibinfo{author}{\bibfnamefont{N.}~\bibnamefont{Katsarakis}},
  \bibinfo{author}{\bibfnamefont{T.}~\bibnamefont{Koschny}},
  \bibinfo{author}{\bibfnamefont{M.}~\bibnamefont{Kafesaki}},
  \bibinfo{author}{\bibfnamefont{E.~N.} \bibnamefont{Economou}},
  \bibnamefont{and} \bibinfo{author}{\bibfnamefont{C.~M.}
  \bibnamefont{Soukoulis}}, \bibinfo{journal}{Applied Physics Letters}
  \textbf{\bibinfo{volume}{84}}, \bibinfo{pages}{2943} (\bibinfo{year}{2004}),
  \urlprefix\url{http://link.aip.org/link/?APL/84/2943/1}.

\bibitem[{\citenamefont{Smith et~al.}(2006)\citenamefont{Smith, Gollub, Mock,
  Padilla, and Schurig}}]{smith_jap_2006}
\bibinfo{author}{\bibfnamefont{D.~R.} \bibnamefont{Smith}},
  \bibinfo{author}{\bibfnamefont{J.}~\bibnamefont{Gollub}},
  \bibinfo{author}{\bibfnamefont{J.~J.} \bibnamefont{Mock}},
  \bibinfo{author}{\bibfnamefont{W.~J.} \bibnamefont{Padilla}},
  \bibnamefont{and} \bibinfo{author}{\bibfnamefont{D.}~\bibnamefont{Schurig}},
  \bibinfo{journal}{Journal of Applied Physics} \textbf{\bibinfo{volume}{100}},
  \bibinfo{eid}{024507} (pages~\bibinfo{numpages}{9}) (\bibinfo{year}{2006}),
  \urlprefix\url{http://link.aip.org/link/?JAP/100/024507/1}.

\bibitem[{\citenamefont{Schneider et~al.}(2009)\citenamefont{Schneider,
  Shuvaev, Engelbrecht, Demokritov, and Pimenov}}]{schneider_prl_2009}
\bibinfo{author}{\bibfnamefont{A.}~\bibnamefont{Schneider}},
  \bibinfo{author}{\bibfnamefont{A.}~\bibnamefont{Shuvaev}},
  \bibinfo{author}{\bibfnamefont{S.}~\bibnamefont{Engelbrecht}},
  \bibinfo{author}{\bibfnamefont{S.~O.} \bibnamefont{Demokritov}},
  \bibnamefont{and} \bibinfo{author}{\bibfnamefont{A.}~\bibnamefont{Pimenov}},
  \bibinfo{journal}{Physical Review Letters} \textbf{\bibinfo{volume}{103}},
  \bibinfo{eid}{103907} (pages~\bibinfo{numpages}{4}) (\bibinfo{year}{2009}),
  \urlprefix\url{http://link.aps.org/abstract/PRL/v103/e103907}.

\bibitem[{\citenamefont{Plum et~al.}(2008)\citenamefont{Plum, Fedotov, and
  Zheludev}}]{plum_apl_2008}
\bibinfo{author}{\bibfnamefont{E.}~\bibnamefont{Plum}},
  \bibinfo{author}{\bibfnamefont{V.~A.} \bibnamefont{Fedotov}},
  \bibnamefont{and} \bibinfo{author}{\bibfnamefont{N.~I.}
  \bibnamefont{Zheludev}}, \bibinfo{journal}{Applied Physics Letters}
  \textbf{\bibinfo{volume}{93}}, \bibinfo{eid}{191911}
  (pages~\bibinfo{numpages}{3}) (\bibinfo{year}{2008}),
  \urlprefix\url{http://link.aip.org/link/?APL/93/191911/1}.

\bibitem[{\citenamefont{Plum et~al.}(2009{\natexlab{b}})\citenamefont{Plum,
  Liu, Fedotov, Chen, Tsai, and Zheludev}}]{plum_prl_2009}
\bibinfo{author}{\bibfnamefont{E.}~\bibnamefont{Plum}},
  \bibinfo{author}{\bibfnamefont{X.-X.} \bibnamefont{Liu}},
  \bibinfo{author}{\bibfnamefont{V.~A.} \bibnamefont{Fedotov}},
  \bibinfo{author}{\bibfnamefont{Y.}~\bibnamefont{Chen}},
  \bibinfo{author}{\bibfnamefont{D.~P.} \bibnamefont{Tsai}}, \bibnamefont{and}
  \bibinfo{author}{\bibfnamefont{N.~I.} \bibnamefont{Zheludev}},
  \bibinfo{journal}{Physical Review Letters} \textbf{\bibinfo{volume}{102}},
  \bibinfo{eid}{113902} (pages~\bibinfo{numpages}{4})
  (\bibinfo{year}{2009}{\natexlab{b}}),
  \urlprefix\url{http://link.aps.org/abstract/PRL/v102/e113902}.

\bibitem[{\citenamefont{Kozlov and Volkov}(1998)}]{kozlov_book}
\bibinfo{author}{\bibfnamefont{G.~V.} \bibnamefont{Kozlov}} \bibnamefont{and}
  \bibinfo{author}{\bibfnamefont{A.~A.} \bibnamefont{Volkov}},
  \emph{\bibinfo{title}{Millimeter and Submillimeter Wave Spectroscopy of
  Solids}} (\bibinfo{publisher}{Springer}, \bibinfo{address}{Berlin},
  \bibinfo{year}{1998}), p.~\bibinfo{pages}{51}.

\bibitem[{\citenamefont{Pimenov et~al.}(2005)\citenamefont{Pimenov, Tachos,
  Rudolf, Loidl, Schrupp, Sing, Claessen, and Brabers}}]{pimenov_prb_2005}
\bibinfo{author}{\bibfnamefont{A.}~\bibnamefont{Pimenov}},
  \bibinfo{author}{\bibfnamefont{S.}~\bibnamefont{Tachos}},
  \bibinfo{author}{\bibfnamefont{T.}~\bibnamefont{Rudolf}},
  \bibinfo{author}{\bibfnamefont{A.}~\bibnamefont{Loidl}},
  \bibinfo{author}{\bibfnamefont{D.}~\bibnamefont{Schrupp}},
  \bibinfo{author}{\bibfnamefont{M.}~\bibnamefont{Sing}},
  \bibinfo{author}{\bibfnamefont{R.}~\bibnamefont{Claessen}}, \bibnamefont{and}
  \bibinfo{author}{\bibfnamefont{V.~A.~M.} \bibnamefont{Brabers}},
  \bibinfo{journal}{Phys. Rev. B} \textbf{\bibinfo{volume}{72}},
  \bibinfo{pages}{035131} (\bibinfo{year}{2005}).

\bibitem[{\citenamefont{Oksanen et~al.}(1990)\citenamefont{Oksanen, Tretiakov,
  and Lindell}}]{oksanen_jewa_1990}
\bibinfo{author}{\bibfnamefont{M.}~\bibnamefont{Oksanen}},
  \bibinfo{author}{\bibfnamefont{S.}~\bibnamefont{Tretiakov}},
  \bibnamefont{and} \bibinfo{author}{\bibfnamefont{I.}~\bibnamefont{Lindell}},
  \bibinfo{journal}{J. Electromagn. Waves Appl.} \textbf{\bibinfo{volume}{4}},
  \bibinfo{pages}{613} (\bibinfo{year}{1990}).

\bibitem[{\citenamefont{Berreman}(1972)}]{berreman_josa_1972}
\bibinfo{author}{\bibfnamefont{D.}~\bibnamefont{Berreman}},
  \bibinfo{journal}{J. Opt. Soc. Am.} \textbf{\bibinfo{volume}{62}},
  \bibinfo{pages}{502} (\bibinfo{year}{1972}), ISSN \bibinfo{issn}{0030-3941}.

\end{thebibliography}

\end{document}